\documentclass[conference]{IEEEtran}
\ifCLASSINFOpdf
\else
\fi
\usepackage{color}
\usepackage{enumerate}
\usepackage[cmex10]{amsmath} 
\usepackage{siunitx}
\usepackage{import} 
\usepackage{enumitem}

\usepackage{indentfirst} 
\usepackage{standalone}
\usepackage[section]{placeins} 
\usepackage{pdflscape}
\usepackage{tabularx}
\usepackage{algorithm} 
\usepackage{amssymb} 
\usepackage{microtype} 
\usepackage{soul} 
\usepackage{mathtools}
\usepackage{multirow}
\usepackage{verbatim}
\usepackage{graphicx}

\usepackage{pgfplots}

\usepackage{subfloat} 
\usepackage{verbatim}

\usepackage[utf8]{inputenc}
\usepackage[T1]{fontenc}
\usepackage{amsmath} 
\usepackage[shellescape,latex]{gmp} 
\usepackage[space]{grffile}
\usepackage{booktabs} 
\usepackage{xfrac}
\usepackage{tabularx}

\usepackage{cite}
\usepackage{adjustbox}

\usepackage{verbatim}
\usepackage{url}

\usetikzlibrary{shapes}

\usepackage{multicol, blindtext}    
\usepackage[]{color}
\usepackage{framed}
\usepackage{alltt}
\usepackage[T1]{fontenc}
\usepackage{array}
\usepackage{colortbl}
\usepackage{booktabs}
\usepackage{multirow}
\usepackage{eurosym}
\usepackage{listings}
\usepackage{ifthen}
\usepackage{pgfplots}
\usepackage{pgfplotstable}
\usepackage{pgfcalendar}
\usepackage{pgfgantt}
\usepackage{tikz}
\usetikzlibrary{shapes.geometric, arrows}

\usepackage{longtable}
\usepackage{amsfonts}
\usepackage{subfigure}
\usepackage{float}

\usepackage{mathrsfs}
\usepackage{algpseudocode} 
\usepackage{textcomp}
\usepackage{xspace}

\makeatletter
\newcommand\fs@betterruled{%
  \def\@fs@pre{\vspace*{5pt}\hrule height.8pt depth0pt \kern2pt}%
  \def\@fs@post{\kern2pt\hrule\relax}%
  \def\@fs@mid{\kern2pt\hrule\kern2pt}%
  \let\@fs@iftopcapt\iftrue}
\floatstyle{betterruled}
\restylefloat{algorithm}
\makeatother

\pgfplotsset{compat=1.16}

\usepackage{fmtcount} 

\usepackage[
  top=0.72in,
  bottom=1in,
  left=0.64in,
  right=0.64in
]{geometry}
\begin{document}

\title{Multi-Agentic AI for Conflict-Aware rApp Policy Orchestration in Open RAN
\vspace{-0.3cm}
}

\author{
\IEEEauthorblockN{Haiyuan Li, Yulei Wu, Dimitra~Simeonidou}
\IEEEauthorblockA{\textit{High Performance Network Group, Smart Internet Lab, Faculty of Engineering, University of Bristol, U.K.}\\
E-mail: \{ocean.h.li, y.l.wu, dimitra.simeonidou\}@bristol.ac.uk}
}

\maketitle

\begin{abstract}
Open Radio Access Network (RAN) enables flexible, AI-driven control of mobile networks through disaggregated, multi-vendor components. In this architecture, xApps handle real-time functions, whereas rApps in the non-real-time controller generate strategic policies. However, current rApp development remains largely manual, brittle, and poorly scalable as xApp diversity proliferates. In this work, we propose a Multi-Agentic AI framework to automate rApp policy generation and orchestration. The architecture integrates three specialized large language model (LLM)-based agents, Perception, Reasoning, and Refinement, supported by retrieval-augmented generation (RAG) and memory-based analogical reasoning. These agents collectively analyze potential conflicts, synthesize intent-aligned control pipelines, and incrementally refine deployment decisions. Experiments across diverse deployment scenarios demonstrate that the proposed system achieves over 70\% improvement in deployment accuracy and 95\% reduction in reasoning cost compared to baseline methods, while maintaining zero-shot generalization to unseen intents. These results establish a scalable and conflict-aware solution for fully autonomous, zero-touch rApp orchestration in Open RAN.

\end{abstract}


\section{Introduction} \label{sec:Introduction}
Open Radio Access Network (Open RAN) has emerged as a transformative, software-defined paradigm that meets the flexibility, responsiveness, and service-aware control demanded by immersive AR/VR, autonomous vehicles, industrial automation, digital-twin systems, and other next-generation mobile applications~\cite{polese2023understanding}. By disaggregating the traditionally monolithic RAN stack into interoperable modules connected through open interfaces, Open RAN allows operators to mix-and-match multi-vendor components, embed intelligence directly into network operations, and realize truly programmable, closed-loop control~\cite{villa2025x5g}. Its layered architecture distributes decision-making across two key control planes: xApps in the Near-Real-Time RAN intelligent controller (Near-RT RIC) execute latency-sensitive functions, such as AI-assisted mobility management, spectrum allocation, and traffic steering on sub-second timescales, while rApps in the Non-Real-Time RIC (Non-RT RIC) generate strategic, long-horizon policies that coordinate underlying heterogeneous xApps~\cite{d2023orchestran}. This separation of policy reasoning from time-critical enforcement provides the modularity and scalability required for dynamic network slicing and heterogeneous quality-of-service (QoS) guarantees.

Despite these advantages, the current development of rApps remains largely closed, manual, and vendor-specific, limiting interoperability and slowing innovation in Open RAN~\cite{marinova2024intelligent}. 
As the number and diversity of xApps grow, each tailored to different vendors or services, the design of rApp policies implemented through various xApp compositions becomes increasingly complex and difficult to scale. Furthermore, a critical challenge lies in the growing risk of coordination conflicts, both intra-layer (xApp–xApp and rApp–rApp) and inter-layer (xApp–rApp), that emerge as control logic becomes increasingly distributed and heterogeneous.
In response, prior research has proposed various approaches to automate rApp policy orchestration and mitigate policy-level conflicts.
Broadly, these can be categorized into (i) Rule-based strategies~\cite{adamczyk2023conflict, del2025pacifista, wadud2025xapp}, such as rollback mechanisms and first-come-first-serve (FCFS) schemes, which resolve policy conflicts using static priorities or heuristics; (ii) Game-theoretic methods~\cite{wadud2024qacm, wadud2023conflict}, which model xApp coordination as strategic interactions among agents, aiming to reach stable or optimal equilibria under resource constraints; and (iii) Classic AI-based solutions, including deep reinforcement learning (DRL)~\cite{cinemre2025xapp, zhang2022team, zafar2025conflict}, which enables multiple xApps to learn coordinated control policies through joint interaction with the network environment, and knowledge distillation~\cite{erdol2024xapp}, which transfers knowledge from multiple xApps into a single model that preserves their original functionality. However, existing solutions exhibit several critical limitations in achieving scalable, automated, and intelligent rApp orchestration; these include:

\begin{itemize}[leftmargin=*]
    \item \textit{Oversimplified conflict modeling:} Current coordination mechanisms primarily address conflict coordination between xApps, while overlooking broader coordination challenges involving both xApps and rApps. 
    \item \textit{Limited generalization and scalability:} Most prior methods are designed for static coordination scenarios and fail to generalize to unseen intents, dynamic xApp configurations, or evolving network environments.
   \item \textit{Lack of automation:} Existing rApp development remains largely manual, requiring expert intervention to interpret intents, define coordination logic, and validate policy safety. This reliance on human-in-the-loop processes limits the openness, automation, and responsiveness promised by the Open RAN paradigm.
\end{itemize}

To overcome these limitations, this paper introduces an Agentic AI framework for scalable and automated rApp policy generation and orchestration in Open RAN. The main contributions are summarized as follows:
\begin{itemize}[leftmargin=*]
    \item \textit{Multi-agent framework:} We propose a collaborative Multi-Agentic AI framework composed of three specialized LLM-based agents, including perception, reasoning, and refinement, that work together to automate rApp policy generation from high-level service intents.

    \item \textit{Context- and memory-aware orchestration:} The framework combines conflict-aware perception, generated based on the current deployment context, with critique-based refinement guided by episodic memory. This integration enables the agent ensemble to avoid known failure patterns, mitigate hallucinations, and achieve stable convergence through monotonic policy improvement across iterations.
    
    \item \textit{Extensive Experiments:} 
    The proposed framework is evaluated across diverse xApp configurations and deployment scenarios. Compared to baseline methods, it achieves over \textit{70\%} improvement in deployment accuracy and reduces reasoning cost by more than \textit{95\%}. These results demonstrate the effectiveness and scalability of our conflict-aware, LLM-driven orchestration framework toward fully autonomous, zero-touch rApp management in Open RAN.
\end{itemize}

\section{xApp coordination in Open RAN: System context and Problem formulation}
\label{sec:formulation}

\begin{figure}[t]
    \centering
    \includegraphics[width=1\linewidth]{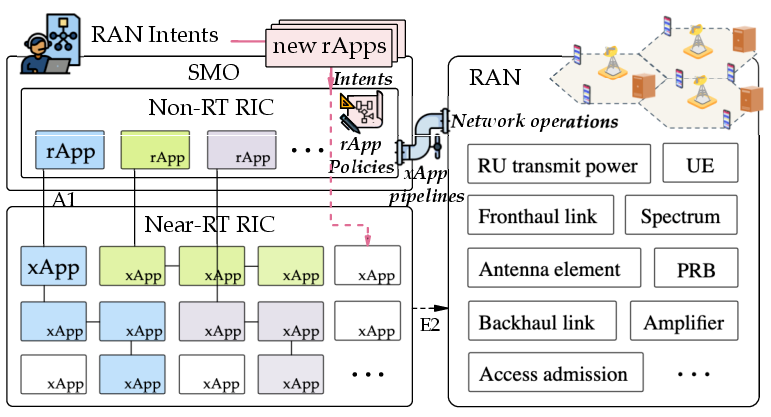}
    \vspace{-0.5cm}
    \caption{An Open RAN scenario where high-level RAN intents are translated into rApp policies, each instantiated as a pipeline of xApps managing real-time network attributes.}
    \vspace{-0.4cm}
    \label{fig:scenario}
\end{figure}

As illustrated in Fig.~\ref{fig:scenario}, we consider a dynamic scenario in which a batch of new RAN intents, denoted $\{I_j\}_{j=1}^m$, is submitted to the service management and orchestration (SMO) by the network operator. Each intent $I_j$ specifies a high-level network objective, such as enhancing mobility robustness, reducing energy consumption, or ensuring bounded latency, and requires the synthesis of a corresponding rApp policy $r_j$ to fulfill it. These rApp policies are constructed by selecting and composing xApps from a heterogeneous function pool $\mathcal{X}$, pre-registered in the Near-RT RIC. Simultaneously, a subset of previously deployed rApps, denoted $\Pi_{\text{cur}}$, may already be active in the Non-RT RIC as a result of earlier orchestration decisions.
Each rApp policy $\pi_j$ can be represented as a directed acyclic graph (DAG) $(\mathcal{X}_j, E_j)$, where $\mathcal{X}_j \subseteq \mathcal{X}$ denotes the selected xApps, and $E_j \subseteq \mathcal{X}_j \times \mathcal{X}_j$ encodes execution dependencies such that an edge $(x_a, x_b) \in E_j$ implies that $x_a$ must precede $x_b$ during execution.

Since the SMO may be required to handle a batch of intents $\{I_k\}_{k=1}^m$ concurrently, a central challenge lies in ensuring that the resulting rApp pipelines are both effective and mutually compatible. Specifically, we identify four major forms of coordination conflicts that can occur among xApps and rApps, including both intra-layer (xApp-xApp, rApp-rApp) and inter-layer (xApp-rApp) interactions:

\begin{itemize}[leftmargin=*, label=-]
    \item \textit{Actuator Contention:} Multiple rApps may simultaneously select and attempt to configure the same xApp instance, producing conflicting or incompatible control directives at the actuator level.
    \item \textit{Parameter Coupling:} Distinct xApps may independently attempt to control the same underlying network parameters (e.g., transmit power, scheduling weights).
    \item \textit{Objective Interference:} rApps may pursue opposing targets for a shared key performance indicator (KPI). In addition, an xApp invoked by one rApp may also degrade system-wide performance metrics that are essential to other rApps. These interactions cause both direct and implicit conflicts at the objective level.
    \item \textit{Vendor-level Interoperability Conflict:} Open interfaces such as A1 and E2 do not eliminate differences across vendor-specific xApps. Semantic mismatches, inconsistent parameter conventions, or proprietary extensions may prevent rApp policies from executing reliably or consistently across deployments.
\end{itemize} 

Overall, the orchestration objective is to synthesize, for each incoming intent $I_j$ in the set $\{I_j\}_{j=1}^m$, an optimal rApp policy $\pi_j'$ that fulfills the corresponding network objective 
in isolation; and to deploy the largest possible subset of these policies $\{\pi_j'\}_{j=1}^m$ such that the union of all active pipelines, including the pre-existing rApps $\Pi_{\text{cur}}$, remains conflict-free.
The objective function can be written as:
\begin{equation}
\max _{\left\{\pi_j\right\}_{j=1}^m}
\left|
\left\{
\pi_j \mid 
\pi_j = \pi_j' \ \land \ 
\pi_j \in \mathcal{V}\left(
\Pi_{\text{cur}} \cup \{\pi_k\}_{k \ne j}
\right)
\right\}
\right|
\label{equ:obj}
\end{equation}
where $\mathcal{V}(\cdot)$ denotes the validity function that evaluates whether a given rApp pipeline $\pi_j$ can be safely deployed in the presence of other active pipelines $\{ \pi_k \}_{k \ne j}
$. Because each intent may be fulfilled through combinatorial compositions of xApps of arbitrary length, the resulting unbounded action space becomes intractable for traditional AI or rule-based methods. Moreover, complex and interdependent coordination conflicts further compound the orchestration challenge.

\section{Multi-Agentic AI for rApp policy generation} \label{sec:agentai}
\begin{figure}
    \centering
    \includegraphics[width=1\linewidth]{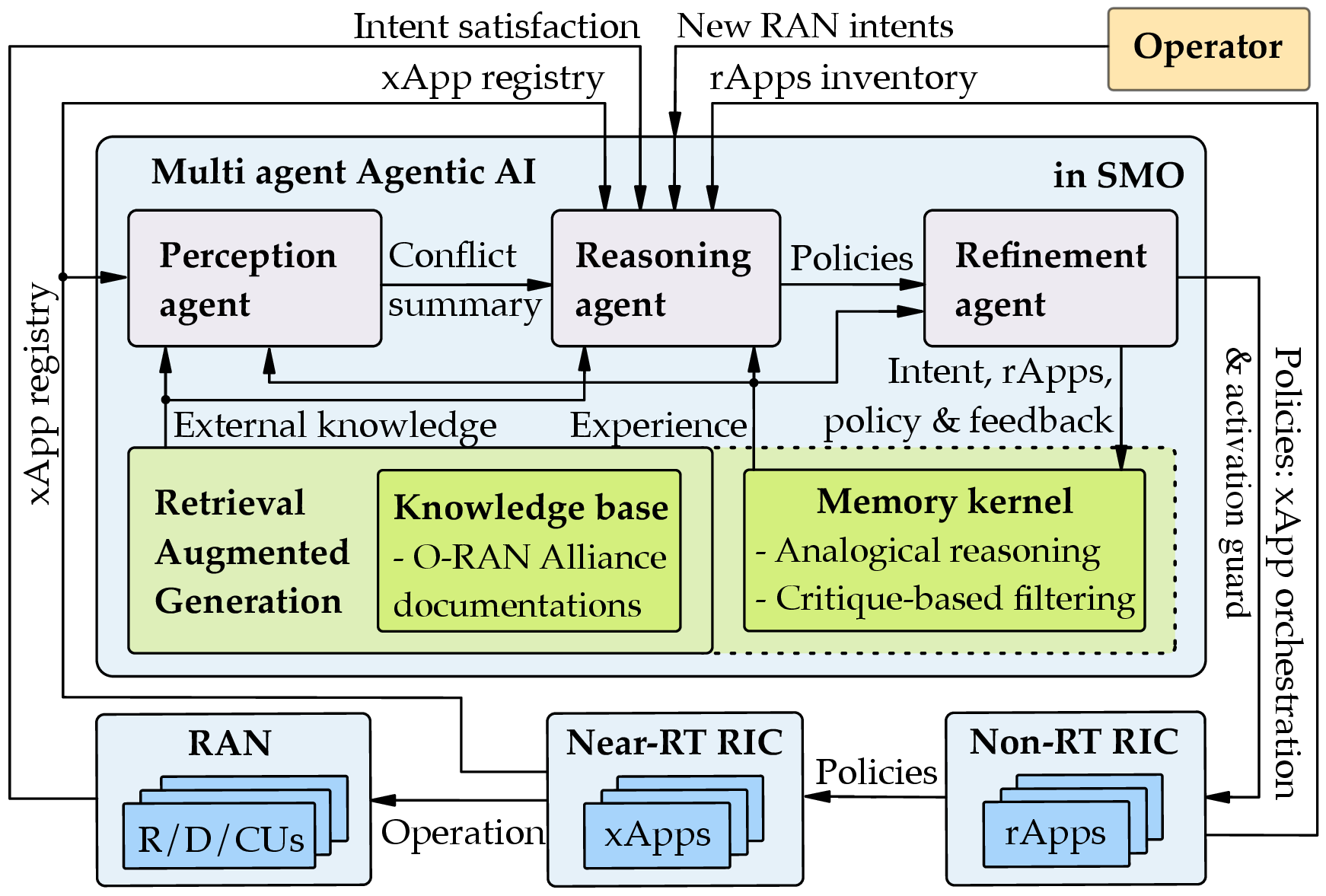}
    \vspace{-0.4cm}
    \caption{The Multi-Agentic AI framework.}
    \vspace{-0.5cm}
    \label{fig:agenticai}
\end{figure}

To address the orchestration and coordination challenges, we propose a cooperative Multi-Agentic AI framework~\cite{busoniu2008comprehensive, olfati2007consensus} for rApp policy generation as shown in Fig.~\ref{fig:agenticai}. This framework comprises three agents driven by large language models (LLMs), which collaborate to support intent translation and conflict-aware pipeline synthesis. The \textit{Perception Agent} first analyzes the current Open RAN environment to identify potential coordination conflicts. Based on this structured conflict representation, the \textit{Reasoning Agent} synthesizes rApp pipelines by selecting and composing xApps that are both effective and safely deployable. Finally, the \textit{Refinement Agent} serves as a structured reviewer, applying incremental corrections to eliminate structural inconsistencies and resolve residual policy-level conflicts.

\subsection{Perception Agent}
The Perception Agent is responsible for constructing a structured representation of potential conflicts present in the current Open RAN environment. 
The agent is initialized with a system prompt that explicitly defines the functional objective and provides a reference taxonomy of conflict types, as described in the previous section. 
Given this system prompt, the module takes three input: (i) the current request $I_j$, representing the service intent or optimization goal to be satisfied (e.g., mobility robustness, latency reduction, energy saving), (ii) the set of available xApps $\mathcal{X}$ in the Near-RT RIC registry, each described by a structured profile capturing its functional role, configurable parameters $\theta_i$, expected inputs and outputs, control interfaces (e.g., O1 configuration management (CM), Near-RT RIC APIs), and known effects on network behavior, and (iii) the currently active rApps and their associated policies $\{ \pi_k \}_{k \neq j}$. The output format is standardized by the system prompt, guiding the agent to produce a structured JSON representation of potential conflicts.

Based on this setup, the agent transforms raw, unstructured inputs into a structured JSON-based representation that captures potential conflicts and semantic relations among xApps. Rather than treating xApps in isolation, it supports reasoning over their joint composition, control dependencies, and cumulative impact on multi-objective constraints and latent interference within the system.

\subsection{Reasoning Agent}
Building on the conflict analysis produced by the Perception Agent, this Reasoning Agent serves as the core decision engine that translates high-level service intents into executable rApp policies, directly optimizing the underlying objective function in Eq.~\ref{equ:obj}. Similar to the Perception Agent, this agent is guided by a system prompt that defines its role in translating intents into deployable policies while maximizing rApp deployment without triggering conflicts. The prompt also includes a structured description of the available xApps in the Near-RT RIC and their corresponding functionalities.

Using this context, it processes the current intent $I_j$, the set of active rApps and their associated policies $\{ \pi_k \}_{k \neq j}$, and the conflict graph produced by the Perception Agent. Based on these inputs, the agent produces a policy $\pi_j = \{ \mathcal{X}_j, \delta_j \}$, where $\mathcal{X}_j \subseteq \mathcal{X}$ denotes the selected xApps that collectively form the executable control pipeline, and $\delta_j$ specifies the deployment conditions under which this pipeline can be safely activated. These conditions account for current network dynamics and potential interactions with existing rApp logic.

\subsection{Knowledge Retrieval and Memory-Augmented Reasoning}
To enhance the accurate conflict analysis and decision quality, as well as ensure consistency with the domain-specific operational norms of the two agents, these agents are enhanced with retrieval-augmented generation (RAG)~\cite{lewis2020retrieval} with an external retrieval mechanism that can access a knowledge base that includes technical documentation from O-RAN Alliance Work Groups. The agents retrieve semantically similar documents with the system prompt and input context and inject them together into the prompt to enrich the LLM input, enabling the agent to reason over protocol constraints, best practices, and domain-specific policies when analyzing potential conflicts and generating intent-aligned control pipelines.

In addition, we incorporate a Memory Kernel that enables the agents to learn from prior decision outcomes. This module maintains an episodic memory buffer $\mathcal{M}$ recording tuples of the form:
\begin{equation}
\left(I_j, \pi_j, \mathcal{C}_j\right)
\end{equation}
where $\mathcal{C}_j$ encodes observed effectiveness of deploying policy $\pi_j$ for intent $I_j$. $\mathcal{M}$ serves as a structured repository that enables temporal reasoning and experience-based policy refinement.

Rather than relying solely on stochastic sampling (e.g., temperature-based exploration)~\cite{renze2024effect}, the agents can reuse prior experience from $\mathcal{M}$ to improve the conflict analysis and decision robustness in two complementary ways. \textit{Analogical reasoning} is applied by retrieving previously successful intent–policy pairs $(I_k, \pi_k)$ from memory. These exemplars are provided to the LLM in a few-shot format to guide the synthesis of candidate pipelines for the current intent $I_j$, enabling the reuse of effective patterns observed under similar network conditions.

\subsection{Refinement Agent}
While memory mechanisms provide the LLM with access to past knowledge, the information retrieved via RAG may introduce contradictions or policy-level conflicts. Standard LLMs often struggle to detect and resolve these inconsistencies or to robustly integrate corrective feedback across turns~\cite{laban2025llms, huang2023large}.
Coupled with the inherent complexity of the task, which requires the model to perform both analytical assessment and high-level reasoning, it is challenging for the two agents to account for all relevant factors in a single pass.
In response, we introduce a final component in the architecture: the Refinement Agent, which calls the LLM in the role of a structured policy reviewer that revises the single pipeline produced by the Reasoning Agent for the current intent. 

It operates with a targeted context comprising: (i) the candidate pipeline to be reviewed, (ii) a textual summary of recurrent failure patterns retrieved from the Memory Kernel, and (iii) a description of the current deployment context. Using this information, the agent performs localized reasoning to correct structural and mechanical defects, such as duplicate xApps, superfluous selections, and stage‑ordering violations, and to bias the candidate away from failure modes previously observed in memory. Unlike RAG mechanisms that retrieve documents based on semantic similarity to the current input, the Refinement Agent queries the memory buffer specifically for past experiences associated with the same intent. This setup enables the agent to mitigate hallucinations produced by the Reasoning Agent and to identify and correct recurring structural errors introduced during policy synthesis. By ensuring that each revision performs no worse than the previous attempts, the agent enforces a monotonic improvement trajectory throughout the iterative process.

Together, these agents form a closed-loop decision-making framework that jointly evaluates the control scope of available xApps and synthesizes conflict-aware policy pipelines that are semantically aligned with target intents, continuously refining orchestration decisions through contextual analysis and experience-driven feedback. Compared with rule-based or conventional AI methods, the LLM-driven architecture exhibits superior generalization to unseen intents and supports variable-length reasoning, enabling seamless adaptation to dynamic xApp configurations and policy constraints without retraining or reconfiguration.

\section{Experiment and results}
To evaluate the effectiveness of our proposed agent-based framework for intent-driven rApp policy generation in Open RAN, we designed an experimental setup that simulates a RAN control environment with heterogeneous xApps and multiple intent-triggered rApps. This section describes the composition of the environment, the architecture of the agentic system used during evaluation, and the comparative results across multiple deployment scenarios.

\subsection{Experimental setup}
\subsubsection{Open RAN environment}
We consider a total of 14 xApps covering diverse functional areas in the RAN control stack. These include (i) an LSTM-based Mobility Predictor that anticipates UE handover targets based on past movement patterns, (ii - iii) two Traffic Steering Agents (A/B variants) with functionally similar logic but different control semantics, designed to emulate vendor-specific incompatibilities, (iv) a Power Saving Controller that scales transmission power or schedules sleep-mode transitions for underutilized cells, balancing energy savings with edge coverage trade-offs, (v) a Spectrum Sharing Optimizer that dynamically allocates physical resource blocks (PRBs) based on real-time interference and occupancy, aiming to improve spectral efficiency, (vi) a Latency-Aware MAC Scheduler that performs low-latency traffic prioritization using real-time CSI and flow deadlines, potentially at the cost of broadband throughput, (vii) a Wireless Anomaly Detector that detects abnormal RAN behavior (e.g., unexpected SINR drops), publishing alarms to upper layers, (viii) a Massive MIMO Beamformer that computes downlink/uplink beamforming weights from channel state information (CSI) to maximize SINR, (ix) an Uplink Power Control Agent that tunes UE-side transmit power to minimize UL interference while maintaining Signal to Noise Ratio (SNR) targets, (x) a Baseband Placement Scheduler which maps baseband function chains to edge/cloud resources under latency and fronthaul constraints, (xi) an Admission Control Manager that makes slice-aware admission decisions in response to dynamic traffic loads and SLA constraints, (xii - xiii) two RAN Slicing Managers (A/B variants) that allocate per-slice resources based on SLA forecasts designed by two different vendors, and (xiv) an URLLC Guard that enforces strict latency bounds for URLLC flows via dynamic preemption and queue prioritization policies.

The experimental workload consists of seven RAN intents, each corresponding to a distinct service-level objective. These include ($1$) "Enhance mobility robustness for high-speed UEs.", ($2$) "Minimise RAN energy consumption during off-peak hours.", ($3$) "Guarantee sub-5 ms E2E latency for factory-automation slice.", ($4$) "Detect and mitigate wireless traffic anomalies in real time.", ($5$) "Maximise video-streaming throughput with limited spectrum.", ($6$) "Guarantee deterministic latency for URLLC under load surges.", and ($7$) "Assure slice isolation with vendor-A slicing in multi-traffic scenarios." To evaluate the performance of the proposed solution under diverse deployment scenarios, we consider settings where a subset of rApps corresponding to predefined intents is already deployed, and a new set of RAN intents arrives as additional requests to be handled in each decision-making instance. For each intent, a corresponding target policy is pre-defined to serve as the ground truth for evaluation of the Agentic AI generated policy and the reference implementation for rApps that are selected to be pre-deployed in the network.

\subsubsection{Agentic AI}
The multi-agent framework described in Section~\ref{sec:agentai} is implemented using LLM-based reasoning modules deployed via OpenAI's GPT-5 API, accessed through the LangChain interface. All agents operate under a structured system, prompts, and support RAG and memory-augmented reasoning. 

Each agent is instantiated as an independent reasoning module with a distinct system prompt and JSON-based I/O schema. The Perception Agent, for example, classifies coordination risks using a fixed taxonomy and outputs a typed JSON object representing actuator-level, parameter-level, vendor-level, and intent-level conflicts. These outputs are consumed by the Reasoning Agent, which generates executable xApp pipelines and deployment conditions using filtered xApp pools and structured conflict maps. The Refinement Agent acts as a post-processing component, validating candidate outputs and applying structured edits when constraint violations or suboptimal deployment plans are detected.

Both the RAG mechanism and the Memory Kernel are supported via a local Chroma vector store, which is pre-loaded with O-RAN documentation and internal domain knowledge. Documents are preprocessed into 500-character segments with 50-character overlaps and embedded using OpenAI’s \textit{text-embedding-3-small} model. At runtime, relevant entries are retrieved via cosine similarity based on the current intent and associated xApp identifiers. During retrieval, the number of documents injected into the agent’s context window is dynamically adjusted according to the current iteration step: the process starts with the top-10 most similar entries and progressively increases the retrieval size across iterations to incorporate more experience-informed context, capped at a maximum of 50 documents.

\begin{figure}[t]
    \centering
    \includegraphics[width=1\linewidth]{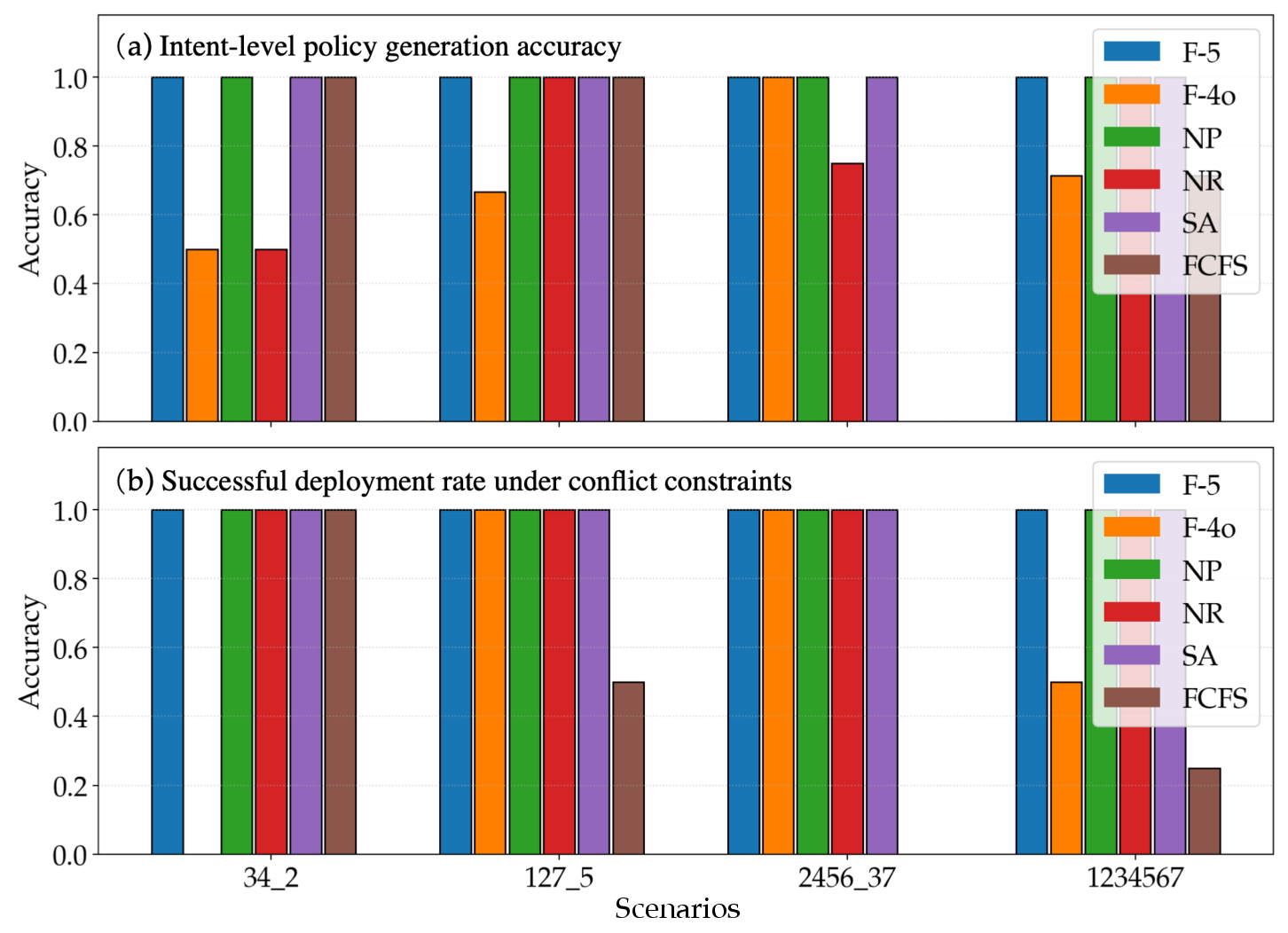}
    \vspace{-0.8cm}
    \caption{Evaluation of policy generation accuracy and conflict-aware deployment success rate (While several baselines achieve comparable accuracy, they may differ significantly in reasoning efficiency (see Fig.~\ref{fig:attempts}).).}
    \vspace{-0.3cm}
    \label{fig:accuracy}
\end{figure}

\subsection{Results}
To evaluate the effectiveness and efficiency of the proposed multi-agent orchestration framework, we assess its performance along two dimensions: 
(i) \textit{solution accuracy}, defined as the ability to synthesize and successfully deploy conflict-free rApp pipelines that fulfill the requested service intents; and 
(ii) \textit{reasoning efficiency}, measured by the number of iterations required to reach such deployment (capped at 50).

Specifically, we design four orchestration scenarios of increasing complexity, each defined by a batch of new RAN intents submitted to the SMO. The goal of the agent is to synthesize, for each intent $I_j$, an optimal xApp pipeline $\pi_j$ that (i) fulfills the intent's service-level objective and (ii) can be safely deployed in conjunction with other pipelines, including pre-existing rApps, without introducing coordination conflicts.
The four scenarios are as follows:
(i) intents $\{3, 4\}$ with rApps previously generated from intent $\{2\}$ already deployed;  
(ii) intents $\{1, 2, 7\}$ with pre-deployed rApps originating from intent $\{5\}$;  
(iii) intents $\{2, 4, 5, 6\}$ with rApps derived from intents $\{3, 7\}$ already present in the system;
(iv) all seven intents $\{1\text{--}7\}$ in an initially empty rApp inventory.
These scenarios are selected to cover a gradient of complexity, from minimal intent-intent interaction (Scenario 1) to densely overlapping multi-intent orchestration (Scenario 4). In each case, the pre-deployed rApp combinations are manually selected to ensure that a ground-truth conflict-free deployment exists, allowing consistent benchmarking.

\begin{figure}[t]
    \centering
    \includegraphics[width=1\linewidth]{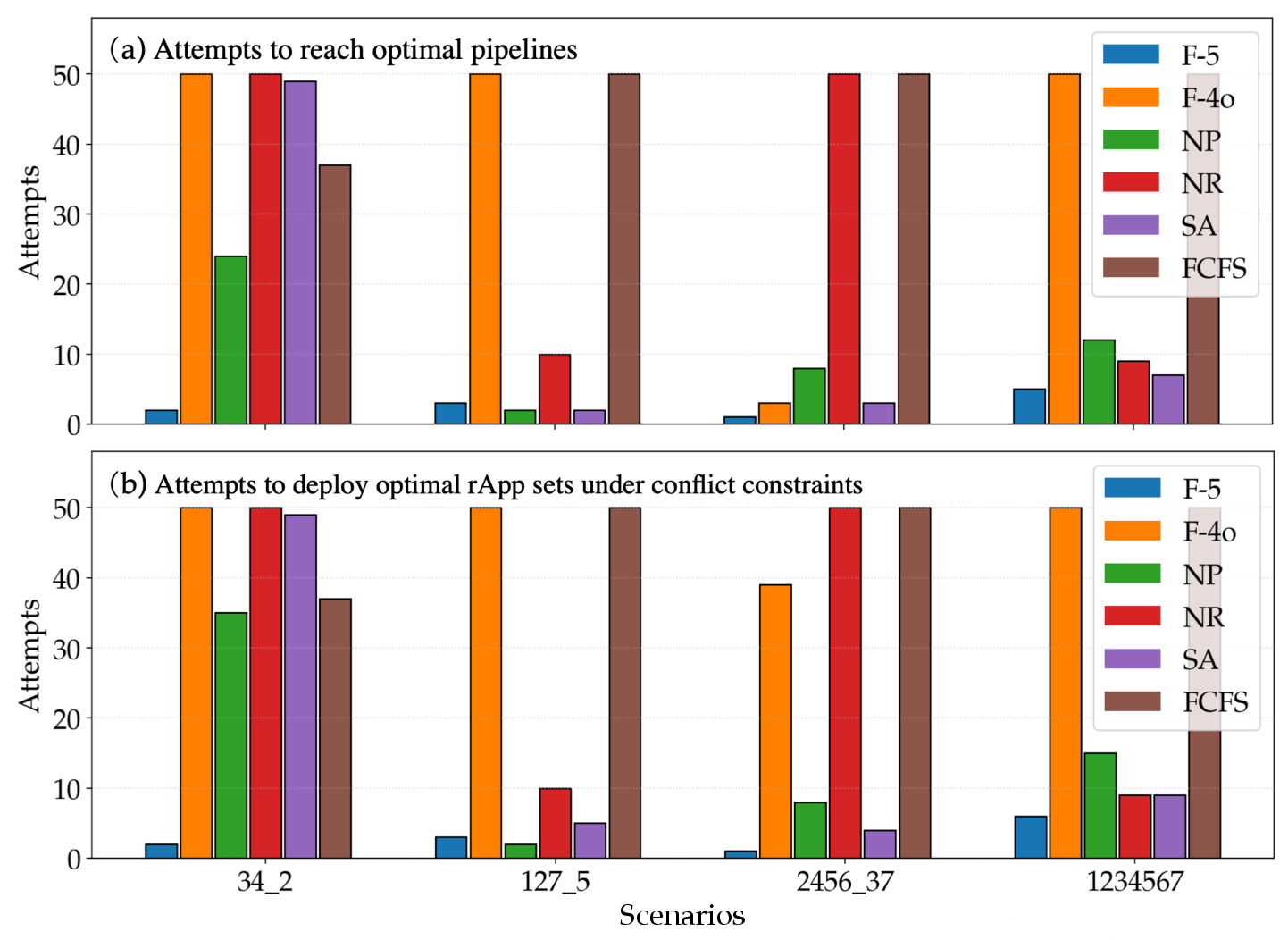}
    \vspace{-0.7cm}
    \caption{Evaluation of orchestration efficiency: attempts required for optimal policy generation and deployment.}
    \vspace{-0.3cm}
    \label{fig:attempts}
\end{figure}
For each scenario, the proposed solution is initialized with a cleared memory buffer and iteratively constructs executable pipelines for the incoming intents. The process continues until it reaches the ground-truth solution, defined as generating the correct pipeline for each request and achieving the maximum number of conflict-free rApp deployments.
To ensure evaluation consistency and limit computational overhead, each scenario allows a maximum of 50 iterations. If convergence is not achieved within this limit, the attempt is considered unsuccessful.
We compare our full solution (\textit{F-5}) against five baselines:
\begin{itemize}[leftmargin=*, label=-]
  \item \textit{Single-Agent (SA):} A single LLM maps the entire input, including xApp descriptions, active rApps, and intent requests, directly to pipeline composition and deployment decisions.
  \item \textit{No Refinement (NR) Agent:} A two-agent version of our framework that omits the final refinement agent responsible for structured critique and error correction.
  \item \textit{No Perception (NP) Agent:} A version without the Perception Agent, removing explicit conflict detection prior to policy generation.
  \item \textit{Full solution based on GPT-4o (F-4o):} The full multi-agent framework using GPT-4o, to assess model sensitivity to language model capacity.
  \item \textit{First-Come-First-Served (FCFS):} A variant where pipelines are generated using the same agent-based synthesis framework, but deployment follows a fixed first-come, first-served strategy. Each optimized pipeline is deployed only if it does not conflict with any previously deployed rApps; otherwise, it is skipped.
\end{itemize}

We first examine the solution accuracy across all scenarios by evaluating both the rate of optimal pipeline synthesis and the success rate of conflict-free deployment. As shown in Fig.~\ref{fig:accuracy}, our full solution (F-5), along with SA and NP, achieves perfect accuracy across all four scenarios. In contrast, NR fails to reach optimality in Scenarios~1 and~3. This reflects the importance of the Refinement Agent in preventing the language model from making elementary or systematic errors that may not be mitigated through prompting alone. In addition, the weakest performance is observed in F-4o and FCFS. The F-4o configuration often fails to synthesize valid pipelines or make effective deployment decisions. This result highlights the importance of sufficient LLM reasoning capability, particularly in handling multi-intent composition and conflict-aware orchestration. FCFS also performs poorly, especially in the last three scenarios where inter-intent conflicts become more prominent. Unlike the other configurations, it lacks the ability to reason over global coordination constraints. Therefore, it is unable to resolve policy-level incompatibilities during deployment, leading to suboptimal outcomes.

We next assess the reasoning efficiency, measured by the number of iterations required to achieve the observed accuracy. 
As shown in Fig.~\ref{fig:attempts}, the proposed F-5 solution consistently converges within five attempts for pipeline synthesis and within six iterations for conflict-free deployment across all scenarios.
In contrast, both SA and NP require significantly more attempts to reach comparable accuracy levels. In Scenario~1, SA requires 49 iterations and NP takes 35, while F-5 completes the same task in only two iterations, yielding reductions of 95.9\% and 94.3\%, respectively. A similar pattern is observed in Scenario~4, where SA and NP take 10 and 15 iterations, compared to just five iterations under F-5. These results represent a 50\% and 66.7\% reduction in reasoning cost.
This performance gap can be attributed to a shared limitation of SA and NP: the lack of a dedicated perception mechanism for early conflict detection. Without explicit constraint information, both methods rely solely on the LLM’s internal reasoning, which leads to inefficient trial-and-error behavior. By contrast, the Perception Agent in F-5 provides structured conflict analysis prior to pipeline generation, thereby offloading reasoning burden from the generation process and enabling faster convergence without sacrificing deployment quality.

\section{Conclusion} \label{sec:Conclusion}
This paper presents a Multi-Agentic AI framework for automated and scalable rApp policy generation in Open RAN. By coupling LLM-driven perception and reasoning agents with episodic memory and retrieval-based augmentation, the system enables conflict-aware pipeline synthesis that generalizes across diverse xApp configurations and service intents. Experimental results demonstrate that memory-guided analogical reasoning significantly improves convergence speed and deployment success rates, especially in multi-intent scenarios with complex coordination constraints. Compared to single-agent, memory-less, and reduced-capability baselines, the proposed architecture consistently delivers higher orchestration quality with minimal manual intervention. These findings highlight the promise of agentic architectures in advancing closed-loop, zero-touch orchestration for programmable Open RAN.

\section*{Acknowledgments}
\small{The authors would like to express their gratitude for the support of the UK-funded project REASON under the Future Open Networks Research Challenge sponsored by the Department of Science, Innovation and Technology (DSIT). This work was also supported by the NVIDIA Academic Grant Program.}


\bibliographystyle{IEEEtran}
\small{\bibliography{bib}}

\end{document}